**Arthur Zakinyan, Yuri Dikansky, Marita Bedzhanyan**

*North-Caucasian Federal University, Institute of Natural Sciences, Stavropol, Russian Federation*

Address correspondence to A. Zakinyan:

phone: +7-8652-57-00-33,

postal address: North-Caucasian Federal University, Institute of Natural Sciences, Department of Theoretical Physics, 1 Pushkin Street, 355009 Stavropol, Russian Federation,

e-mail: zakinyan.a.r@mail.ru

**Acknowledgments.** This work was supported by the grant of the President of Russian Federation No. MK-6053.2012.2.

# Electrical Properties of Chain Microstructure Magnetic Emulsions in Magnetic Field

Running head: Electrical properties of magnetic emulsions

**Abstract.** The work deals with the experimental study of the emulsion whose dispersion medium is a magnetic fluid while the disperse phase is formed by a glycerin-water mixture. It is demonstrated that under effect of a magnetic field chain aggregates form from the disperse phase drops. Such emulsion microstructure change affects its macroscopic properties. The emulsion dielectric permeability and specific electrical conductivity have been measured. It is demonstrated that under the effect of relatively weak external magnetic fields (~ 1 kA/m) the emulsion electrical parameters may change several fold. The work theoretically analyzes the discovered regularities of the emulsion electrical properties.

## 1. INTRODUCTION

The magnetic emulsions are disperse systems consisting of two liquid phases; one of them is a magnetic fluid. A magnetic fluid is a stable colloidal dispersion of magnetic nanoparticles (~ 10 nm) in a specific carrier fluid. To obtain a magnetic emulsion a magnetic fluid is mixed with a fluid medium insoluble in it. Here, a magnetic fluid may form both the disperse phase and the dispersion medium of the emulsion. The sizes of the resulting disperse phase drops varies between 0.1 to 100 µm, these are several orders greater than those of the magnetic fluid magnetic nanoparticles; hence, the latter may be in most cases considered as a continuous fluid magnetizable medium. The idea to create the magnetic emulsions [1] has originated from the possibility of their advantageous application as a magnetically sensitive medium. The presence of the expressed magnetic properties makes these emulsions unique objects of study and makes the difference between them and the classic emulsions.

Up to today, several works dealing to any extent with the magnetic emulsions physical properties studies have been known. Thus, [2] has revealed that the uniformly distributed spherical drops in magnetic field interact with the formation of the chain aggregates whose orientation is in line with the external field direction. The processes of the chain aggregates formation in magnetic emulsions have been studied in [3–6], as well.

In [7, 8], the optical properties of the magnetic emulsions whose disperse phase is represented by the magnetic fluid drops have been studied. They demonstrate the results of measurement of emulsion transparency depending upon the external magnetic field direction and strength. They describe the formation of the chain aggregates of the drops and dense structures of column type that are, as well, formed by the magnetic fluid drops under effect of magnetic field. The magnetic emulsions rheological properties linked with the chain aggregates in magnetic field formation have been studied in [9]. The specificities of the magnetic properties of the magnetic emulsions whose disperse phases is a magnetic fluid have been studied in [10]. It has studied the concentration and temperature dependences of the emulsion magnetic susceptibility and the character of its magnetization in constant magnetic field.

Some results of study of the magnetic emulsions where the dispersion medium is a magnetic fluid and the disperse phase consists of the glycerin drops are given in [11]. It mentions the formation of the chain aggregates from the emulsion drops under effect of external magnetic field on such system. It points out the occurrence of the anisotropic light scattering due to the chain aggregates. It should be noted that the earlier [12–16] had studied the structure formation occurring in a magnetic fluid under effect of magnetic field when nonmagnetic microparticles are placed into it. These had studied, essentially, the microstructure, and, partly, certain macroscopic properties of the system of solid nonmagnetic microparticles in magnetic fluid under effect of magnetic field.

In [17–19], some specificities of the optical, magnetic, and electrical properties of the magnetic emulsions due to the disperse phase drops deformation under effect of external magnetic fields are described. Such deformation under effect of the relatively weak external magnetic fields is possible due to the smallness of the interfacial tension at the disperse phase – dispersion medium boundary of the emulsions studied in [17–19]. At a great enough interfacial tension the disperse phase drops do not deform; in this event, a chain microstructure forms.

The said works note that even the effect of comparatively weak fields may provoke, in magnetic emulsions, the structure formation affecting the medium macroscopic properties. It should be noted that the exposure to a magnetic field does not, practically, provoke the changes of the macroscopic properties of the initial medium; in any event, such change is extremely small and may be neglected when studying the magnetic emulsions properties. Thus, the studies of the magnetic emulsions macroscopic properties are of great interest due to the possibility of their efficient control by exposing to external force fields.

This work studies the specificities of the macroscopic electrical properties of the magnetic emulsions provoked by the formation of the chain aggregates from the disperse phase drops under effect of external magnetic field. Such structure formation must lead to peculiar specificities of the electrical properties of such media; especially, the anisotropy of the emulsions electrical properties and dependence of the latter upon the magnetic field strength are to be expected.

## 2. EXPERIMENTAL AND RESULTS

For the study object the magnetic emulsion was used whose dispersion medium is a magnetic fluid while the disperse phase is formed by a 1:1 glycerin-water mixture. The emulsion was prepared of a polyethylsiloxane-based magnetite magnetic fluid. The initial magnetic permeability of the magnetic fluid was $\mu \cong 4$; its dielectric permeability was $\varepsilon_e \cong 5$, and the specific electrical conductivity was $\lambda_e \cong 5\cdot 10^{-7}$ S/m. The interfacial tension at the magnetic fluid – glycerin boundary was ~ 0,01 N/m. The glycerin-water mixture was selected as the disperse phase substance because such mixture has the comparatively great dielectric permeability ($\varepsilon_i \cong 60$) and specific electrical conductivity ($\lambda_i \cong 5\cdot 10^{-5}$ S/m) compared with a magnetic fluid dispersion medium. Such emulsion may be classified as "water-in-oil"; due to the great dielectric permeability and specific conductivity of the disperse phase compared with the dispersion medium, a significant change of the macroscopic electrical properties of such emulsion during structure formation under effect of external fields can be expected. The emulsion was prepared by dispersing some glycerin-water mixture in magnetic fluid with an electromechanical mixer. The mean size of the emulsion disperse phase drops was 5 µm. When preparing the emulsion, no emulsifiers were used: these could have provoked a significant alteration of the emulsion electrical properties; hence, complications in interpreting the results.

First, the structural properties of the glycerin drops in magnetic fluid synthesized emulsions under effect of external magnetic field were studied. For this, an emulsion specimen was placed in a constant uniform magnetic field created by Helmholtz coils. The structural transformations in emulsion were observed through an optical microscope. When such a medium is exposed to a magnetic field, the nonmagnetic drops enclosed in a fluid magnetizable medium may be treated as the diamagnetic particles ("magnetic holes") whose magnetic moments are in opposition to the field. The magnetic moments interaction provokes the nonmagnetic drops aggregation into the chain structures oriented along the field intensity vector. Figure 1 shows, for example, a chain structure formed in a magnetic emulsion layer under effect of uniform constant magnetic field. It should be noted that the experimental magnetic field intensities did not provoke any visible deformation of the emulsion disperse phase drops. The emulsion structural state observations have also shown that when the volume concentration grows significantly ($>70\%$) the emulsion passes to a gel-like state with a strongly decreased fluidity.

The specific electrical conductivity $\lambda$ and dielectric permeability $\varepsilon$ of the emulsions were studied using the bridge method. Figure 2 shows the experimental setup layout. To determine the dielectric permeability and specific conductivity, the emulsion under study was placed into a standard cell *1* with platinum coated electrodes to reduce the measurement error due to the electrodes polarization. The distance between electrodes was 2 mm. Then the cell capacity and its active conductivity were measured using the parallel equivalent circuit. The measurements used the 1 kHz measurement signal E7–15 type digital alternating current bridge *3*. The sought dielectric permeability $\varepsilon$, and specific conductivity $\lambda$ were calculated from the expression:

$$\varepsilon = \frac{C}{C_0}, \quad \lambda = \frac{G}{A}, \qquad (1)$$

where $C_0, C$ are the capacities of the empty and emulsion filled cells; $G$ the active conductivity of the emulsion filled cell; $A$ the constant of cell predetermined using the standard 0.01 aqueous solution of potassium chloride.

To study the effect of the magnetic field on the measured values, the emulsion filled cell was placed into the constant uniform magnetic field created by Helmholtz coils *2*. The measurements were made for both the parallel and perpendicular orientations of the external magnetic and measuring electrical fields. It should be noted that the experimental measuring electrical field was weak enough and did not have effect on the structural state of the emulsion under study, i.e., the structure in emulsion formed only under the effect of the applied external magnetic field.

The dependences of the dielectric permeability and specific electrical conductivity of magnetic emulsion upon the disperse phase concentration and upon the magnetic field strength and direction were studied, as well.

Figure 3 shows the dependences of the emulsion specific electrical conductivity upon the disperse phase volume ratio ($\varphi$), with and without exposure to the magnetic field codirectional with the electrical (measuring) one. It is seen that these dependences are of nonlinear character: the glycerin share growth provokes the emulsion conductivity growth. The effect of the magnetic field changes the character of the concentration dependences: these are above the dependence without field and are of a different curvature. The character of the concentration dependences (see Figure 4) of the emulsion dielectric permeability is analogous.

It turned out that the electrical conductivity and dielectric permeability of all the studied specimens of emulsions depends heavily upon the applied magnetic field strength and direction. Figure 5 shows the dependence of the emulsion specific electrical conductivity upon the external magnetic field strength when parallel and perpendicular to the electrical field for the three glycerin volume concentrations. The dependences shown in Figure 6 of the dielectric permeability of the emulsion exposed to the magnetic field are of similar character. These Figures show that under effect of the magnetic field whose direction coincides with that of the electrical measuring field, the specific conductivity and dielectric permeability grow several fold compared with the initial ones. At the orthogonally related directions of the fields the emulsion permeability and conductivity show only an insignificant decrease.

The changes of the dielectric permeability and specific conductivity in parallel fields differ quantitatively depending on the emulsion disperse phase concentration. Figure 7 shows, for example, the dependences of the specific electrical conductivity relative changes due to the magnetic field effect upon the disperse phase concentration for different external magnetic field strengths. Similar dependences exist for the dielectric permeability, as well. The Figure shows that the greatest relative change of the conductivity under effect of field occurs in the emulsions whose disperse phase volume concentration lies within 20% to 30% where the changes of the permeability and conductivity are most significant. The dielectric permeability and specific conductivity may change several fold; this is a significant result in view of possibility of practical application of the effects under study.

It should be noted that due to the difference of densities of the disperse phase and dispersion medium of the emulsions under study they stratify with time. The emulsion stratification process may be inhibited by introducing a specific surface-active substance; however, as noted above, no additional emulsion stabilization was applied. The study has shown that the stratification process is relatively slow and the resulting change of the emulsion properties may take place only several hours after its preparation. Meanwhile, measuring of the emulsion electrical properties takes about several minutes; hence, the change of properties due to stratification may be neglected.

### 3. ANALYSIS OF RESULTS

To analyze the specificities of the magnetic emulsions electrical properties due to the formation in them of the chain aggregates in magnetic field, the processes of structure formation occurring in such medium should be examined. For this, the formation of chain aggregates in magnetic emulsion under effect of magnetic field was numerically modeled. The modeling used the Monte Carlo method with Metropolis algorithm. This algorithm as applied to a study of formation of chain aggregates in monodisperse suspension of nonmagnetic spherical microparticles in magnetic fluid under effect of magnetic field was first applied and described in detail in [20]. The algorithm consists in the following. First, the particles of the suspension under study are initially arranged in random manner. Then the particles coordinates are changed in random manner; at any such random migration the distance of migration of a particle cannot exceed five diameters of such particle. The total energies of the dipole-dipole interaction of all the particles in system before their random migration and in the new state are calculated. The energy of interaction of every pair of particles is defined from the formula

$$W = \frac{\mu_0 \mu_e}{4\pi} \frac{m_1 m_2 \left(1 - 3\cos^2 \theta\right)}{r^3}, \qquad (2)$$

where

$$m = \frac{3(\mu_i - \mu_e)}{\mu_i + 2\mu_e} HV$$

is the disperse phase drop magnetic moment; $r$ the distance between drops; $V$ the drop volume. In the case under study: $\mu_i = 1$, $\mu_e = 4$.

If the new state total energy is smaller than the initial state one, the new state is fixed, then the new random rearrangement of particles in system occurs. If the energy is greater than in the initial state, the random number $0 \leq x \leq 1$ is generated, which is then compared to $y = \exp(-\Delta E/kT)$; here $\Delta E$ is the change of the system energy at the particle migration. If $y > x$, the new system state is fixed, as well; if not, the particle returns to the previous position. This procedure repeats for as long as the change of energy of every particle becomes commensurable with $kT$.

The stable configuration of the suspension particles resulting from the described procedure corresponds to the system energy minimum.

The chain aggregates in magnetic emulsion were modeled similarly to that described in [20], taking into account the system polydispersity. To account for the particles distribution by sizes, the normal distribution was used with the 2.5 µm particles mean radius and 1.2 dispersion. Besides, only two-dimensional problem (two-dimensional system of drops) was considered.

Figure 8 shows, for example, the results of modeling of the chain aggregates in magnetic emulsion under effect of a uniform constant magnetic field directed horizontally along the Figure plane. The scale in Figure 8 is that the length of side of every image corresponds to 500 µm. Figure 8 shows 1000 particles of disperse phase; this corresponds to the emulsion volume concentration about 6%. Figures 1 and 8 show that when the chain aggregates form, the disperse phase particles by no means always line up strictly along a straight line. For example, individual or several drops join the aggregate on sides; the interaction between aggregates may distort their shape and they cease to be rectilinear.

The results of numerical modeling served to determine the dependence of the mean number of drops in aggregate upon the emulsion disperse phase volume concentration. The study went up only to the 10% concentration; above, the modeling was difficult due to the significant increase of the calculation time. Figure 9 shows the resulting dependence of the number $n$ of particles in chain aggregate upon the emulsion concentration at the magnetic field strength $H =$ 4.2 kA/m. The points in Figure 9 correspond to the numerical modeling data; the solid line shows the approximation of these data with cubic polynomial. This approximating dependence was further on used to analyze the revealed specificities of the magnetic emulsion electrical properties.

Let us examine the effect of the chain aggregates formation in magnetic emulsion on its electrical properties. It should be noted in advance that the electrical field frequency at which the emulsion electrical parameters dispersion may occur can be found from the expression [21]:

$$f_0 = \frac{2\lambda_e + \lambda_i + \varphi(\lambda_e - \lambda_i)}{2\pi\varepsilon_0(2\varepsilon_e + \varepsilon_i + \varphi(\varepsilon_e - \varepsilon_i))}. \tag{3}$$

When substituting in (3) the relevant parameters existing in the described above experiments, we get the dispersion frequency at different concentrations within the range 9 to 13 kHz. The emulsion specific electrical conductivity and dielectric permeability were measured at the 1 kHz measuring field frequency, This means that in the described above experiments the low frequency limit values were measured. In this case, the emulsion dielectric permeability $\varepsilon$ and specific conductivity $\lambda$ without magnetic field may be determined in accordance with the Hanai theory [21] with the following expressions:

$$\varepsilon\left(\frac{3}{\lambda - \lambda_i} - \frac{1}{\lambda}\right) = 3\left(\frac{\varepsilon_e - \varepsilon_i}{\lambda_e - \lambda_i} + \frac{\varepsilon_i}{\lambda - \lambda_i}\right) - \frac{\varepsilon_e}{\lambda_e}, \tag{4}$$

$$\frac{\lambda - \lambda_i}{\lambda_e - \lambda_i}\left(\frac{\lambda_e}{\lambda}\right)^{1/3} = 1 - \varphi. \tag{5}$$

Figure 10 shows the calculation of the emulsion dielectric permeability without magnetic field, according to expression (4); for comparison it also gives the relevant experimental points. A good agreement is seen between the experimental data and calculation from formula (4). This Figure also shows that the emulsion dielectric permeability at high concentrations becomes significantly higher than the permeability of any phase; this means a significant contribution of the drops migration polarization into the dielectric permeability of the emulsion under study.

Figure 11 shows the calculation of the emulsion specific electrical conductivity without magnetic field, according to expression (5); for comparison it also gives the relevant experimental points. A good agreement is seen between the experimental data and calculation from formula (5).

Let us consider the electrical properties of the emulsion exposed to magnetic field and due to the chain aggregates formation processes. Suppose that a chain aggregate may be approximated with a prolate ellipsoid of revolution; then the effective medium approximation may be used to describe the macroscopic electrical properties of magnetic emulsion with chain aggregates. As the effective macroscopic electrical conductivity of emulsion at low frequencies of electrical field is, according to the Hanai theory [21], defined by only the conductivity of the emulsion components, the conductivity of a suspension of ellipsoidal particles can be easily calculated by correspondingly modifying the equation (5).

The dielectric permeability situation is more complicated: the low frequency dielectric permeability is, as it has been noted above, significantly affected by the emulsion drops migration polarization. With all this, a coherent theory of polarization of the sufficiently concentrated suspension of ellipsoidal particles taking into account the free charges migration at low frequencies is, up to today, inexistent; moreover, it is quite difficult to develop. In addition, the adequacy of substitution of the chain aggregates with ellipsoidal particles is, in this event, quite doubtful. Hence, the further analysis will deal only with the emulsion specific electrical conductivity.

In the anisotropic effective medium approximation [18, 22] the emulsion specific conductivity will be determined by the following set of equations, of the Bruggeman equation type:

$$\frac{\lambda_\| - \lambda_i}{\lambda_e - \lambda_i}\left(\frac{\lambda_e}{\lambda_\|}\right)^{N_\|} = 1 - \varphi, \qquad (6)$$

$$\frac{\lambda_\perp - \lambda_i}{\lambda_e - \lambda_i}\left(\frac{\lambda_e}{\lambda_\perp}\right)^{N_\perp} = 1 - \varphi. \qquad (7)$$

Here

$$N_\| = \frac{1 - e(\gamma')^2}{2e(\gamma')^3}\left(\ln\frac{1 + e(\gamma')}{1 - e(\gamma')} - 2e(\gamma')\right); \quad N_\perp = \frac{1 - N_\|}{2},$$

where $e(\gamma') = \sqrt{1 - 1/\gamma'^2}$ is the ellipsoidal particle eccentricity; $\gamma' = \gamma\sqrt{\lambda_\perp/\lambda_\|}$; $\gamma$ is the ellipsoid semiaxes ratio. We will consider a chain aggregate in form of sequence of identical drops organized along one straight line. This idealized representation of the aggregate is not fully compliant with the above experimental observation and numerical modeling data; hence, the additional error in the results of analysis of the emulsion electrical properties. Substituting the chain aggregate with the equivalent prolate ellipsoid of revolution we find that this ellipsoid semiaxes ratio $\gamma$ is, obviously, equal to the number of drops in the chain $n$. To determine $n$ let us use the approximation of the numerical modeling results shown in Figure 9. Let us consider that this approximating dependence may be adequately extrapolated at least to the 15% concentration of emulsion, as shown in Figure 9. Then the specific electrical conductivity of the magnetic emulsion exposed to magnetic field may be calculated within the up to 15% concentrations range.

The numerical analysis of the set of equations (6), (7) taking into account the relevant function for $n$ finds the dependence of the magnetic emulsion specific electrical conductivity upon concentration. Figure 12 shows the so calculated dependences of the specific conductivity upon the emulsion disperse phase volume ratio. Curve *1* in Figure shows the theoretical dependence obtained without magnetic field; curve *2* shows the dependence calculated for the $H = 4.2$ kA/m magnetic field parallel to the electrical measuring field. This Figure also shows the relevant experimental points. Figure 12 demonstrates a good enough agreement between the experimental data and calculation results.

Figure 13 shows the compared experimental and theoretical results of the relative change of the conductivity under effect of the $H = 4.2$ kA/m magnetic field parallel to the electrical measuring field, depending on the emulsion concentration. This Figure also shows in this case a good agreement between the experimental data and calculation results.

It should be noted that at high concentrations of the emulsion, the aggregation process becomes more complex and the formation of aggregates in magnetic field cannot anymore be represented as the drops lined up in individual chains. The experiments and numerical modeling have, instead, shown that complex structures form as a result of aggregation of several chains and individual drops. This complicates significantly the theoretical analysis of the magnetic emulsions electrical properties patterns and requires approaches different from the described above.

## 3. CONCLUSION

Hence, this work has studied the new disperse system: magnetodielectric emulsion based on a magnetic fluid and glycerin-water mixture. The formation of the chain aggregates in such media under effect of magnetic field has been demonstrated. The macroscopic electrical properties have been studied. The effect of the chain aggregates formation on the emulsion properties has been shown. These phenomena have been analyzed and theoretically interpreted. Based on the studies, the conclusion can be made that the electrical properties of the synthesized and studied magnetic emulsions depend heavily upon the effect of magnetic field. This means that they are possible to practically apply as magnetically controlled media.

**Figures Captions**

Fig. 1. Magnetic emulsion with glycerin as disperse phase under effect of $H = 4.2$ kA/m uniform constant magnetic field directed horizontally along the Figure plane.

Fig. 2. Experimental setup to study the electrical properties of magnetic emulsions: *1* – electrical measuring cell filled with emulsion under study; *2* – Helmholtz coils; *3* – measuring bridge.

Fig. 3. Concentration dependences of specific electrical conductivity of glycerin emulsion in magnetic fluid under effect of magnetic field codirectional with electrical (measuring) field: *1*: $H = 0$; *2*: $H = 1.8$ kA/m; *3*: $H = 4.2$ kA/m. Lines are the approximation of experimental data.

Fig. 4. Concentration dependences of dielectric permeability of magnetic emulsion under effect of magnetic field codirectional with electrical (measuring) field. Notations: same as Figure 3.

Fig. 5. Experimental dependence of specific electrical conductivity of glycerin emulsion in magnetic fluid upon magnetic field strength. Dark color points: magnetic field strength parallel to electrical field strength; light color points: magnetic field strength perpendicular to electrical field strength. *1*: $\varphi = 0.3$, *2*: $\varphi = 0.2$, *3*: $\varphi = 0.1$. Lines are the approximation of experimental data.

Fig. 6. Dependence of dielectric permeability of magnetic emulsion upon magnetic field strength. Dark color points: magnetic field strength parallel to electrical field strength; light color points: magnetic field strength perpendicular to electrical field strength. *1*: $\varphi = 0.6$; *2*: $\varphi = 0.3$; *3*: $\varphi = 0.1$. Lines are the approximation of experimental data.

Fig. 7. Dependence of relative change of specific electrical conductivity of magnetic emulsion upon concentration of disperse phase at different magnetic field strengths: *1*: $H = 0.9$ kA/m; *2*: $H = 1.8$ kA/m; *3*: $H = 3.7$ kA/m. Lines are the approximation of experimental data.

Fig. 8. Results of numerical modeling of formation of chain aggregates in magnetic emulsion under effect of a uniform constant magnetic field (explained in text).

Fig. 9. Results of numerical modeling of dependence of number of particles in chain aggregate upon emulsion disperse phase volume concentration.

Fig. 10. Dependence of dielectric permeability of magnetic emulsion upon disperse phase volume concentration without magnetic field. Solid line: theoretical dependence calculated according to expression (4); points: experimental.

Fig. 11. Dependence of specific electrical conductivity of magnetic emulsion upon disperse phase volume concentration without magnetic field. Solid line: theoretical dependence calculated according to expression (5); points: experimental.

Fig. 12. Dependence of specific electrical conductivity of emulsion upon disperse phase volume concentration. Curve *1*: theoretical dependence at $H = 0$.

Curve *2*: theoretical dependence at $H = 4.2$ kA/m. Points show the relevant experimental data.

Fig. 13. Relative change of emulsion specific electrical conductivity under effect of $H = 4.2$ kA/m magnetic field depending on disperse phase concentration. Solid line: theoretical dependence; points: experimental.

FIG. 1.

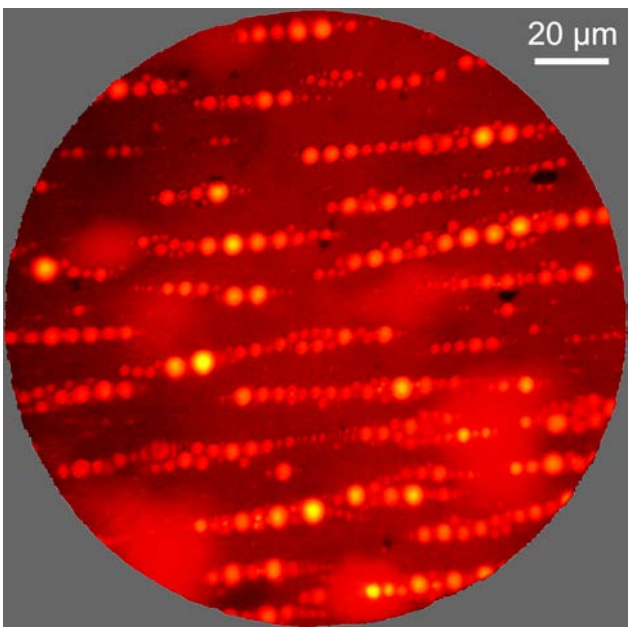

FIG. 2.

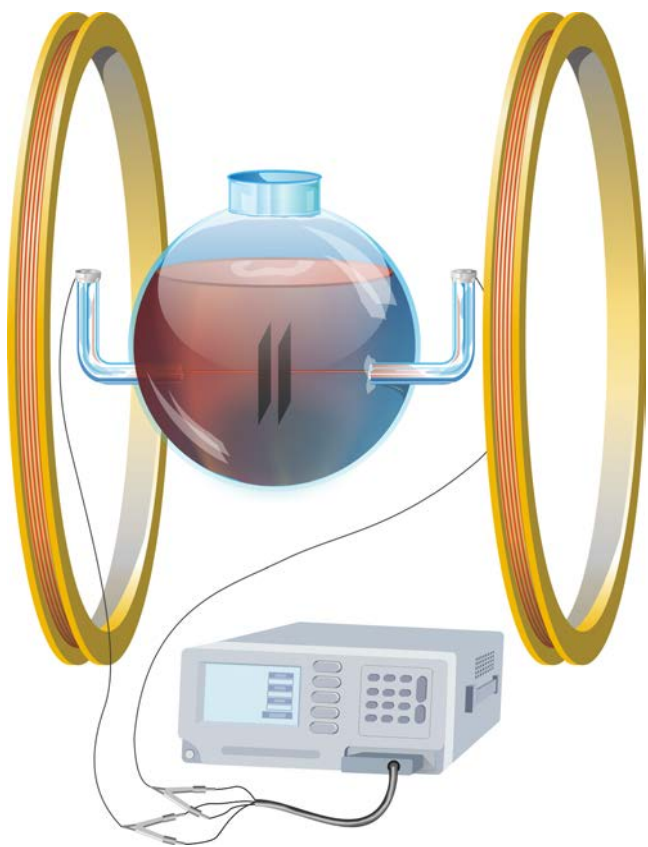

FIG. 3.

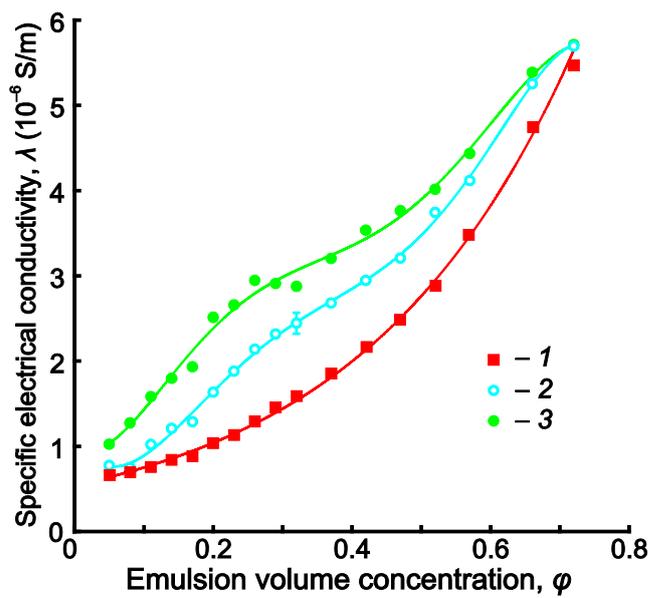

FIG. 4.

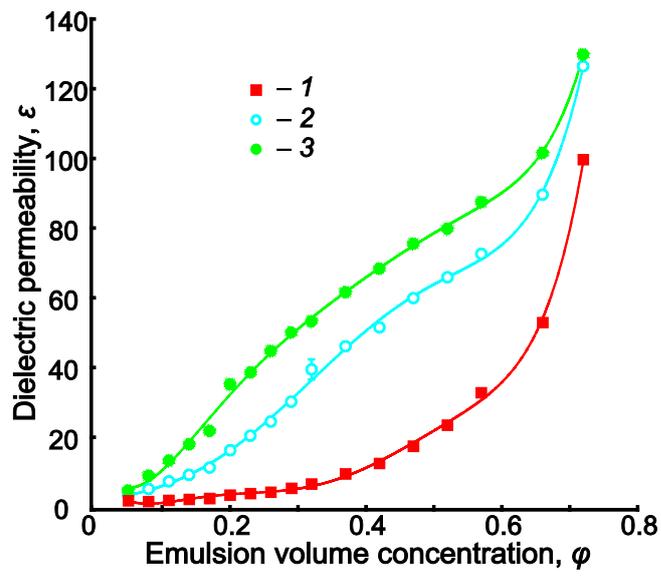

FIG. 5.

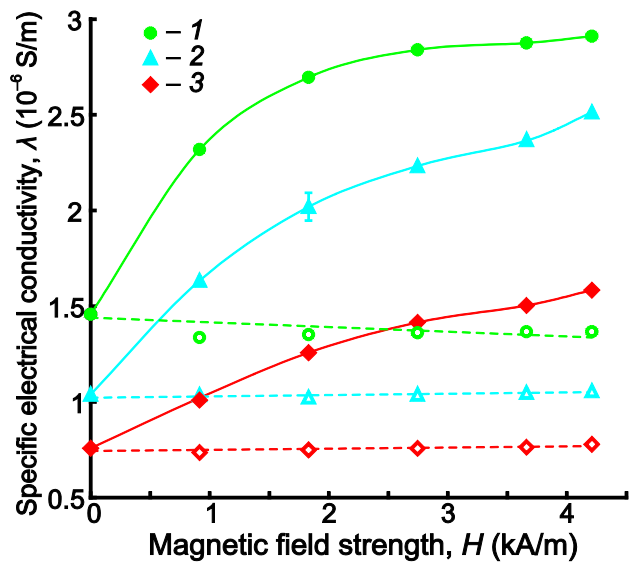

FIG. 6.

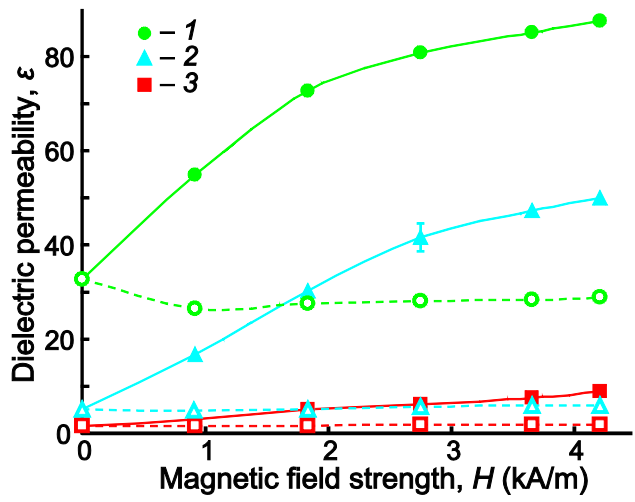

FIG. 7.

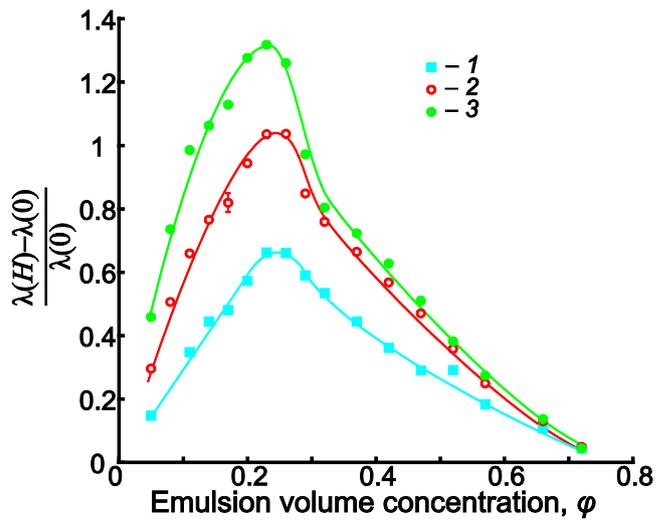

FIG. 8.

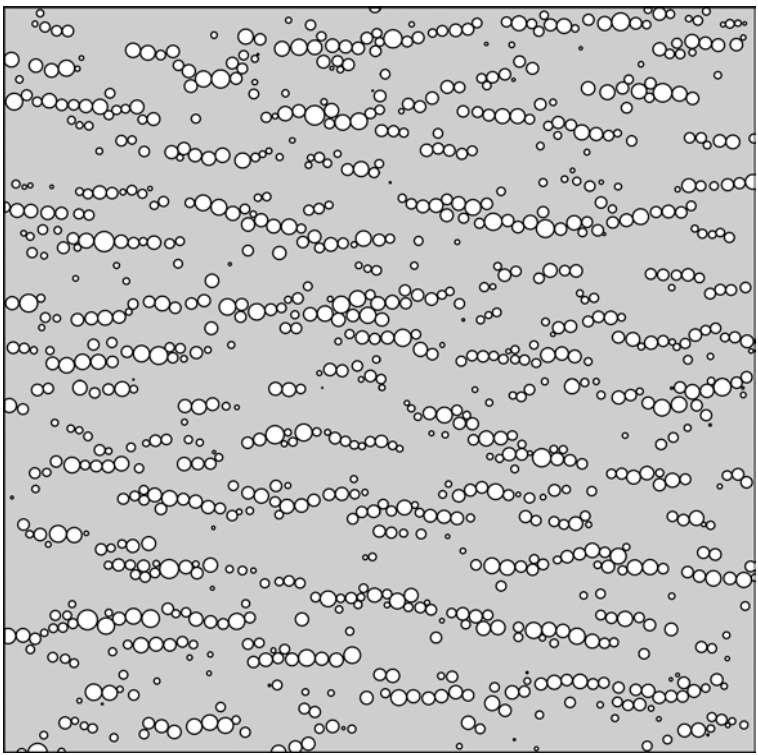

FIG. 9.

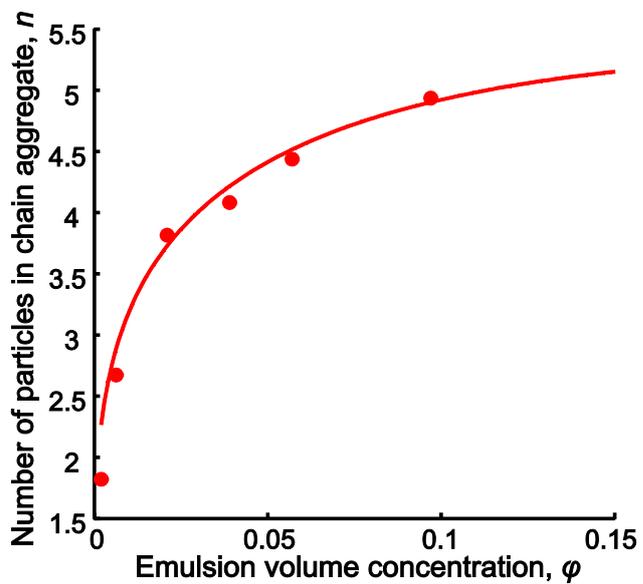

FIG. 10.

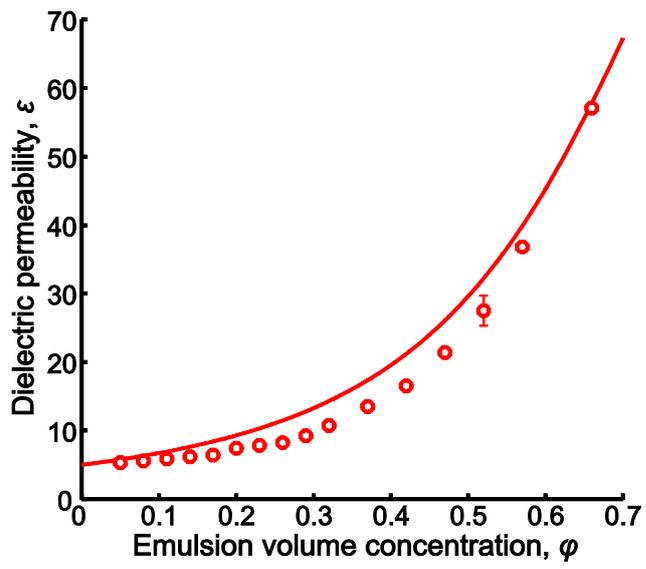

FIG. 11.

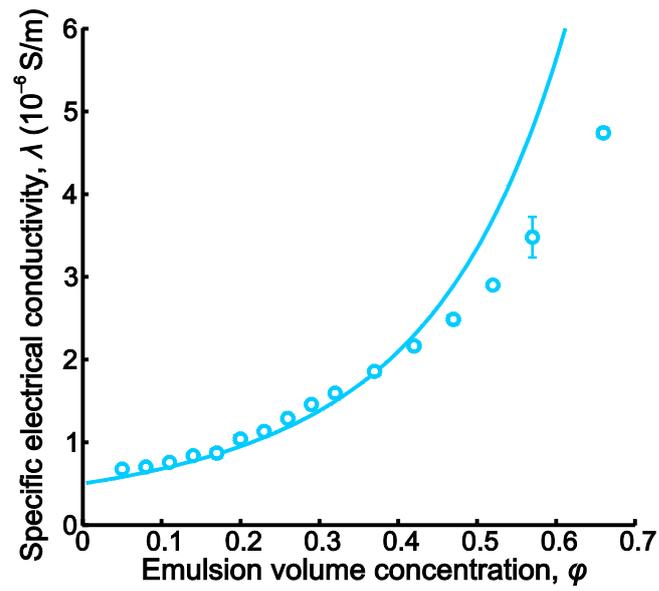

FIG. 12.

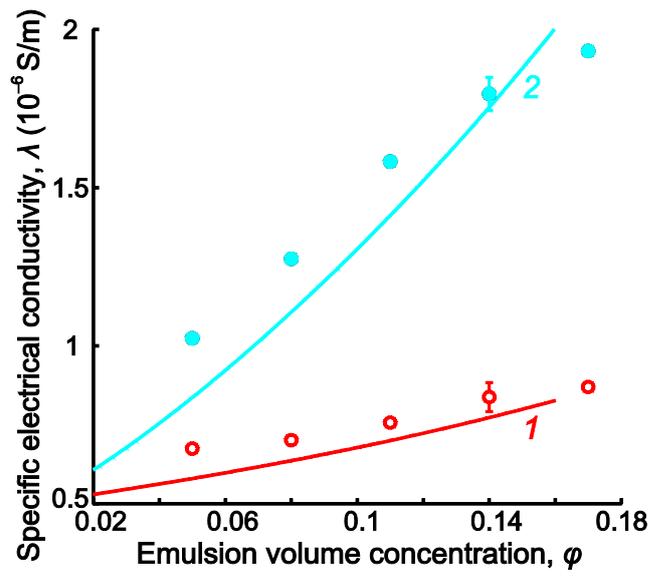

Fig. 13.

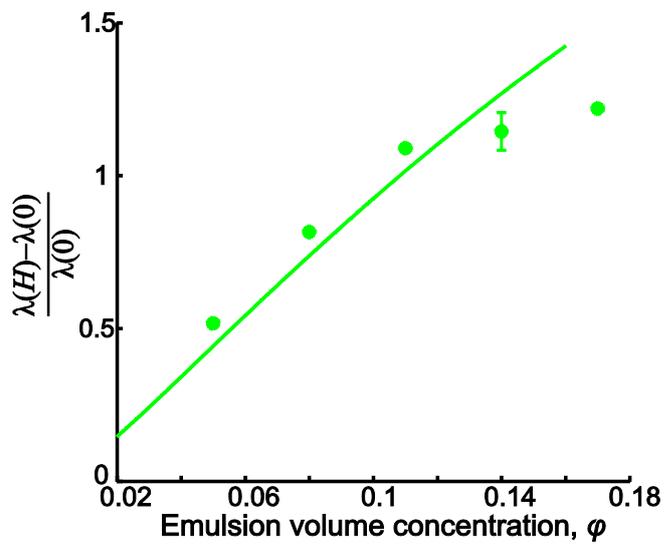